\documentclass[twocolumn]{aastex62}
\usepackage[section]{placeins}
\usepackage{longtable}
\usepackage{hyperref}
\usepackage{url}
\usepackage{xspace}

\newcommand{\WISE}{\textit{WISE}\xspace}

\newcommand{\STScI}{\affiliation{Space Telescope Science Institute, 3700 San Martin Drive, Baltimore, MD 21218, USA}}
\newcommand{\JHU}{\affiliation{Department of Physics and Astronomy, Johns Hopkins University, 3400 North Charles Street, Baltimore, MD 21218, USA}}

\newcommand{\IPACMR}{\affiliation{IPAC, Mail Code 100-22, Caltech, 1200 E.\ California Blvd., Pasadena, CA 91125}}

\newcommand{\CfA}{\affiliation{Center for Astrophysics \textbar{} Harvard \& Smithsonian, 60 Garden Street, Cambridge, MA 02138-1516, USA}}

\newcommand{\DARK}{\affiliation{DARK, Niels Bohr Institute, University of Copenhagen, Jagtvej 128, 2200 Copenhagen, Denmark}}
\newcommand{\UGent}{\affiliation{Sterrenkundig Observatorium, Ghent University, Krijgslaan 281 - S9, B-9000 Gent, Belgium}}
\newcommand{\USzeged}{\affiliation{Department of Experimental Physics, Institute of Physics, University of Szeged, D{\'o}m t{\'e}r 9, 6720 Szeged, Hungary}}
\newcommand{\MTAELTE}{\affiliation{MTA-ELTE Lend\"ulet ``Momentum'' Milky Way Research Group, Szent Imre H. st. 112, 9700 Szombathely, Hungary}}
\newcommand{\Purdue}{\affiliation{Purdue University, Department of Physics and Astronomy, 525 Northwestern Ave, West Lafayette, IN 47907, USA}}
\newcommand{\IDSI}{\affiliation{Integrative Data Science Initiative, Purdue University, West Lafayette, IN 47907, USA}}
\newcommand{\Princeton}{\affiliation{3 Princeton University, 4 Ivy Lane, Princeton, NJ 08544, USA}}
\newcommand{\NARIT}{\affiliation{National Astronomical Research Institute of Thailand, 260 Moo 4,
Donkaew, Maerim, Chiang Mai 50180, Thailand}}
\newcommand{\Virginia}{\affiliation{Department of Physics, Virginia Tech, Blacksburg, VA 24061, USA}}
\shorttitle{SN\,2017gci}
\shortauthors{Gomez et al.}

\begin{document}

\title{Constraining Dust Formation in the Superluminous Supernova 2017gci with JWST Observations}

\correspondingauthor{Sebastian Gomez}
\email{sgomez@stsci.edu}

\author[0000-0001-6395-6702]{Sebastian Gomez}
\STScI

\author[0000-0001-7380-3144]{Tea Temim}
\Princeton

\author[0000-0003-2238-1572]{Ori Fox}
\STScI

\author[0000-0002-5814-4061]{V. Ashley Villar}
\CfA

\author[0000-0002-9301-5302]{Melissa Shahbandeh}
\JHU\STScI

\author[0000-0002-5221-7557]{Chris Ashall}
\Virginia

\author[0000-0001-5754-4007]{Jacob E. Jencson}
\IPACMR

\author[0000-0001-5710-8395]{Danial Langeroodi}
\DARK

\author[0000-0001-9419-6355]{Ilse De Looze}
\UGent

\author[0000-0002-0763-3885]{Dan Milisavljevic}
\Purdue\IDSI

\author[0000-0002-2361-7201]{Justin Pierel}
\STScI

\author[0000-0002-4410-5387]{Armin Rest}
\JHU\STScI

\author[0000-0003-4610-1117]{Tam\'as Szalai}
\USzeged\MTAELTE

\author[0000-0002-1481-4676]{Samaporn Tinyanont}
\NARIT

\begin{abstract}
We present JWST/MIRI observations of the Type I superluminous supernova (SLSN) 2017gci taken over 2000 rest-frame days after the supernova (SN) exploded, which represent the latest phase images taken of any known SLSN. We find that archival \WISE detections of SN\,2017gci taken 70 to 200 days after explosion are most likely explained by an IR dust echo from a $\sim 3 \times 10^{-4}$ M$_\odot$ shell of pre-existing dust, as opposed to freshly-formed dust. New JWST observations reveal IR emission in the field of SN\,2017gci, which we determine is most likely dominated by the host galaxy of the SN, based on the expected flux of the galaxy and the measurable separation between said emission and the location of the SN. Based on models for IR emission of carbonate dust, we place a $3\sigma$ upper limit of $0.83$ M$_\odot$ of dust formed in SN\,2017gci, with a lowest $1\sigma$ limit of $0.44$ M$_\odot$. Infrared (IR) detections of other SLSNe have suggested that SLSNe could be among the most efficient dust producers in the universe. Our results suggest that SLSNe do not necessarily form more dust than other types of SNe, but instead might have a more accelerated dust formation process. More IR observations of a larger sample of SLSNe will be required to determine how efficient dust production is in SLSNe.
\end{abstract}


\section{Introduction} \label{sec:intro}

The large amounts of dust found in high-redshift galaxies at $z \gtrsim 6$ with dust-to-stellar mass ratios around 0.01 to 0.1 \citep{Michalowski10} suggest the need for effective dust formation mechanisms (e.g., \citealt{Liu19, Donevski20, Inami22}). While high-mass asymptotic giant branch (AGB) stars have traditionally been considered major sources of dust, they alone cannot account for all the dust observed in the early universe \citep{Gall11, Vijayan19}. Core-collapse SNe (CCSNe) have also been suggested to be major contributors to the formation of dust \citep{Morgan03, Maiolino04, De17}, with some young core-collapse supernova remnants containing dust masses exceeding 0.3~M$_\odot$ \citep[e.g.][]{Matsuura2015, Temim2017, DeLooze2017}. However, direct mass measurements of dust formation from supernova (SN) light curves have led to a wide range of results. Infrared observations from the \textit{Spitzer} telescope have found masses ranging from $\simeq10^{-5}-10^{-2}$~M$_\odot$ (e.g., \citealt{Sugerman06, Meikle07, Kotak09, Andrews10, Fabbri11, Szalai13}). More recently, JWST/MIRI observations have been used to derive more accurate estimates of the dust mass formed in some SNe. \cite{Shahbandeh23} found at least $\gtrsim 0.014$ M$_\odot$ and $\gtrsim 4.0 \times 10^{-4}$ M$_\odot$ of dust formed in the Type IIP SNe 2004et and 2017eaw, respectively. In contrast, observations of younger SNe at earlier epochs have produced lower mass estimates or more stringent limits. For example, \cite{Hosseinzadeh23} found JWST/MIRI detections of the Type IIb SN 2021afdx to be most consistent with an IR echo from $\sim 4 \times 10^{-4}$~M$_\odot$ of pre-existing dust, and \cite{Shahbandeh24} presented spectral observations of the Type IIP SN 2022acko and found an upper limit on the C/O mass of $< 10^{-8}$ M$_\odot$.

Type I superluminous supernovae (SLSNe), are a class of stripped-envelope CCSNe characterized by a lack of hydrogen in their spectra and high luminosities, up to 100 times more luminous than ``normal" Type Ib/c CCSNe \citep{Chomiuk11, Quimby11, Gal-Yam12, Gal-Yam19}. The conditions that lead a star to produce a SLSN as opposed to a normal Type Ib/c SN are still under debate (e.g., \citealt{Kozyreva15, Yan17, Nicholl17_mosfit, Blanchard21_16inl, Gomez22_LSN, Chen23b}). Nevertheless, the low metallicities observed in the environments of SLSNe (e.g., \citealt{Lunnan14, Angus16, Orum20, Hsu23}) appear to be a key element needed for their progenitors to retain enough angular momentum to produce a millisecond magnetar central engine that can reproduce their light curves (e.g., \citealt{Kasen10, Inserra13}).

Recent studies of SLSNe observed 200-600 days post-explosion suggested that SLSNe could produce $\sim 10 - 100$ times more dust than other types of SNe at similar epochs \citep{Chen21, Sun22}. Infrared observations of the SLSN\,2018bsz extending to $4.5\ \mu$m showed evidence for $\sim 5 \times 10^{-2}$ M$_\odot$ of dust formed 500 days after explosion \citep{Chen21}, and the SLSN\,2020wnt showed evidence of carbon monoxide emission consistent with $\sim 10^{-4}$ M$_\odot$ of dust \citep{Tinyanont23}. \cite{Sun22} presented \WISE observations of a sample of 10 SLSNe and measured the amount of dust formed in two of them (PS15br and SN\,2017ens) to be $\gtrsim 10^{-2}$~M$_\odot$. IR emission from the remaining SNe in that sample, including SN\,2017gci, was found to be consistent with either freshly-formed dust or an IR echo. A robust conclusion proved challenging given that only two bands of \WISE photometry were used, W1 at 3.3~$\mu$m and W2 at 4.6~$\mu$m, which are not red enough to probe the wavelengths above $\sim 5\ \mu$m where dust emission is expected to peak.

The apparent dust production increase seen in some SLSNe could be attributed to several factors. Besides the proposed idea that SLSNe produce more dust than other types of SNe, they may form dust more rapidly, leading to a larger inferred dust mass at early times, without necessarily implying a larger total dust yield (e.g., \citealt{Sarangi13, Gall14}). Alternatively, the rising IR emission observed at early times in some SNe could be due to a lowering of the ejecta opacity, rather than an increase in the dust mass being formed \citep{Dwek19}. The latter scenario appears unlikely given that SLSNe typically have longer diffusion timescales than normal Type Ib/c SNe \citep{Nicholl15}. Here, we aim to constrain the total amount of dust formed in the SLSN\,2017gci to test whether SLSNe show a higher total amount of dust yield than other types of SNe, or if they instead form a comparable amount of dust more rapidly.

SN\,2017gci is a SLSN originally classified by \cite{Lyman17_17gci}. A detailed study from \cite{Fiore21} found that the peak of the light curve of SN\,2017gci is best explained by a magnetar central engine and that an observed late-time flattening is most consistent with being powered by interaction with the circumstellar medium (CSM). Population synthesis models from \cite{Stevance21} found SN\,2017gci to be most consistent with a 30~M$_\odot$ binary system progenitor with an increased mass loss rate via strong stellar winds immediately before death. We use JWST/MIRI observations of SN\,2017gci at over 2000 days after explosion in the $7.7 - 21.0\ \mu$m range to infer the amount of dust formed in the SN as part of JWST program ID 3921 \citep{Fox23}.

This paper is structured as follows: in \S\ref{sec:observations} we present the description of new and archival observations, in \S\ref{sec:methods} and \S\ref{sec:models} we describe our methods and models, respectively, and discuss their implications in \S\ref{sec:discussion}. We conclude in \S\ref{sec:conclusion}. In this work, we adopt the redshift measurement to SN\,2017gci from \cite{Fiore21} of $z = 0.0873$, which corresponds to a luminosity distance of $d_L = 411$ Mpc, where we assume a flat $\Lambda$CDM cosmology based on the Planck 2015 results with \mbox{$H_{0} = 67.8$ km s$^{-1}$ Mpc$^{-1}$} and $\Omega_{m} = 0.308$ \citep{Planck15}.

\section{Observations}\label{sec:observations}

\begin{deluxetable}{cccccc}
    \tablecaption{Log of Observations \label{tab:photometry}}
    \tablewidth{0pt}
    \tablehead{
        \colhead{MJD} &  \colhead{Phase} & \colhead{Mag} & \colhead{Error} & \colhead{Filter} & \colhead{Telescope} \\
                      &  \colhead{Days}  & \colhead{AB}  &                 &                  &                     }
    \startdata
    \multicolumn{6}{c}{SN\,2017gci} \\
    \hline
    58035.5 &   69.1  & 19.84    & 0.25    & W1     & \WISE  \\
    58035.5 &   69.1  & 19.46    & 0.33    & W2     & \WISE  \\
    58191.5 &  212.6  & 20.15    & 0.32    & W1     & \WISE  \\
    58191.5 &  212.6  & 19.12    & 0.23    & W2     & \WISE  \\
    58401.2 &  405.4  & $>$20.16 & \nodata & W1     & \WISE  \\
    58401.2 &  405.4  & $>$19.50 & \nodata & W2     & \WISE  \\
    60191.0 & 2251.7  & $>$25.28 & \nodata & F770W  & JWST  \\
    60191.0 & 2251.7  & $>$24.36 & \nodata & F1000W & JWST  \\
    60191.0 & 2251.7  & $>$22.92 & \nodata & F1500W & JWST  \\
    60191.0 & 2251.7  & $>$21.30 & \nodata & F2100W & JWST  \\
    \hline
    \multicolumn{6}{c}{Host Galaxy} \\
    \hline
    60191.2 & 2052.2  & 24.06    & 0.27    & F336W  & HST   \\
    \nodata & \nodata & 22.83    & 0.14    & g      & PS1   \\
    \nodata & \nodata & 22.28    & 0.14    & r      & PS1   \\
    \nodata & \nodata & 21.99    & 0.08    & i      & PS1   \\
    \nodata & \nodata & 21.86    & 0.16    & z      & PS1   \\
    \nodata & \nodata & 21.74    & 0.29    & y      & PS1   \\
    60412.0 & 2255.4  & 21.96    & 0.07    & i      & DECam \\
    60412.0 & 2255.4  & 21.90    & 0.07    & z      & DECam \\
    60191.0 & 2251.7  & 22.45    & 0.04    & F770W  & JWST  \\
    60191.0 & 2251.7  & 22.57    & 0.09    & F1000W & JWST  \\
    60191.0 & 2251.7  & 21.62    & 0.09    & F1500W & JWST  \\
    60191.0 & 2251.7  & 21.08    & 0.14    & F2100W & JWST  \\
    \enddata
    \tablecomments{Archival and new photometry used in this work. All magnitudes are quoted in the AB system and not corrected for Galactic extinction. The \WISE observations are taken directly from \cite{Sun22}.}
\end{deluxetable}

\begin{figure*}
	\begin{center}
		\includegraphics[width=\textwidth]{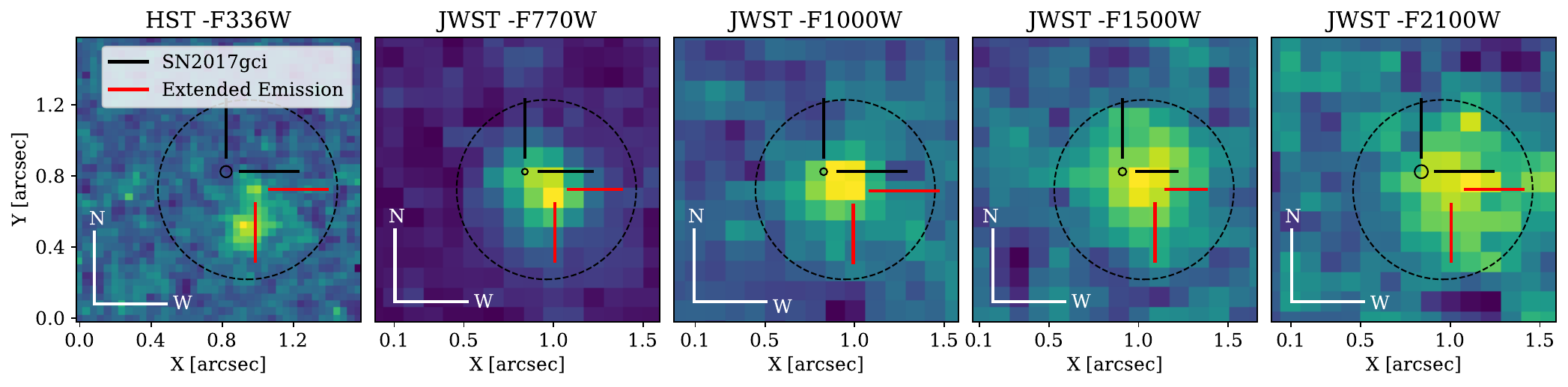}
		\caption{Images of SN\,2017gci from HST and JWST. The HST F336W image is assumed to have no contribution from the SN and represents only its host galaxy. The small solid black circle shows the location of the SN based on its \textit{Gaia} coordinates. The size of the circle represents the astrometric uncertainty of each image, less than 1 pixel for all images. The red cross-hairs point to a source of extended emission seen in the HST image. The dashed circle shows a 0.5\arcsec aperture used to perform photometry on these images. We find the location of the JWST emission to be most consistent with the source of extended emission, as opposed to the location of the SN. \label{fig:target}}
	\end{center}
\end{figure*}

In this section we describe new and archival observations of the field of SN\,2017gci. We correct all photometry for Milky Way extinction with an $E(B-V) = 0.1105$ obtained from the dust maps of \cite{Schlafly11}, assuming $R_V = 3.1$. We use the {\tt Astropy} \citep{astropy} implementation of the \cite{Gordon23} extinction law, which provides accurate correction estimates from the UV to the IR from 912 \AA\ to 32 $\mu m$ \citep{Gordon09, Fitzpatrick19, Gordon21, Decleir22}. All quoted magnitudes are given in AB magnitudes, with a summary of measurements listed in Table~\ref{tab:photometry}. Throughout this paper phases are quoted in rest-frame days from explosion assuming an explosion date of MJD = 57960.42, or 2017 Jul 26, derived in \cite{Gomez24}.

\subsection{JWST/MIRI Imaging}\label{sec:JWST}

We obtained observations of SN\,2017gci as part of JWST Survey Program ID 3921 \citep{Fox23}, a program in the process of obtaining late-time IR observations of a wide range of SNe types, aiming to quantify the amount of dust formed in these. Images were taken with the Mid-Infrared Instrument (MIRI; \citealt{Bouchet15,Rieke15,Ressler15}) with filters F770W, F1000W, F1500W, and F2100W on 2024 April 8, which corresponds to a phase of 2250 days. The images were processed using version 1.13.3 of the JWST Calibration Pipeline with version 11.17.14 of the Calibration Reference Data System (CRDS).

Observations were taken with a four-point dither, which we re-process using the JWST/HST Alignment Tool ({\tt JHAT}, \citealt{Rest2023}) to align each frame to the latest DR3 version of the \textit{Gaia} source catalog \citep{Gaia16, Gaia23}. We perform aperture photometry on the JWST images with an aperture radius of 0.5\arcsec using {\tt space\_phot}\footnote{\url{https://github.com/jpierel14/space_phot}}, and include the latest encircled energy corrections present in the v$1.12.5$ JWST pipeline \citep{bushouse_jwst_2022}. Additionally, we measure $3\sigma$ upper limits of the brightest PSF that can be fit precisely at the location of the \textit{Gaia} coordinates of the SN using {\tt space\_phot} to determine the limits on the total mass of freshly-formed dust.

\subsection{Archival HST Imaging}\label{sec:HST}

SN\,2017gci was observed as part of HST program ID 17181 \citep{Lyman22}, a snapshot proposal that observed a sample of SLSNe with the goal of studying their host galaxies. A single image was taken on 2023 Sep 4 with a three-point dither using WFC3 with the F336W filter. We re-process the individual dithers in the same way as the JWST images using {\tt JHAT} to align the images to \textit{Gaia} sources and perform aperture photometry with a radius of 0.5\arcsec using {\tt space\_phot}.

SLSNe tend to decline rapidly in the UV, completely fading after a few hundred days (e.g., \citealt{Nicholl17_mosfit, Chen23b}). Therefore, it is safe to assume this HST observation taken at a phase of 2052.2 days represents only the host galaxy and has no remaining SN flux.

\subsection{Archival \WISE Imaging}\label{sec:WISE}

SN\,2017gci was detected with \WISE in the W1 and W2 bands as part of the NEOWISE Reactivation Mission \citep{Wright10,Mainzer14} on 2017 Oct 9 and 2018 Mar 14, with a series of subsequent non-detections. We obtain these observations directly from \cite{Sun22} with the only distinction that we convert the Vega magnitudes reported in that work into AB magnitudes.

\subsection{Optical Imaging}\label{sec:optical}

We download pre-explosion archival Pan-STARRS1 3PI Survey (PS1/$3\pi$) images of the field of SN\,2017gci in $grizy$ bands \citep{Chambers16}. Additionally, we obtained $z$ and $y$-band images on 2024 Apr 12 using the Dark Energy Camera (DECam), which were reduced by the standard National Optical-Infrared Astronomy Research Laboratory (NOIRLab) community pipeline \citep{Valdes14}. We assume the DECam images contain no statistically significant supernova light given that they were taken over 2000 days after explosion. This assumption appears robust given the close match to the PS1/$3\pi$ measured host galaxy magnitudes. We perform aperture photometry on all optical images using a 1 \arcsec radius to encompass the full host galaxy and account for the larger ground-based PSF.

\subsection{\textit{Gaia} Detections}\label{sec:gaia}

SN\,2017gci was detected by the \textit{Gaia} Science Alerts (GSA; \citealt{Wyrzykowski16}) soon after explosion at coordinates R.A.$=101.68761$ deg and decl.$=-27.24885$ deg. We use these observations solely with the purpose of having an accurate measurement of the SN location.

\section{Methods}\label{sec:methods}

\subsection{Alignment}\label{sec:alignment}

We align the JWST and HST observations to a common \textit{Gaia} reference frame to verify whether the JWST detections are more consistent with the location of SN\,2017gci or the location of its host galaxy. We use {\tt JHAT} \citep{Rest2023} to re-project the individual dithers of all JWST and HST observations in all filters and align them to sources found in the \textit{Gaia} DR3 source catalog. We align the JWST images to 9 \textit{Gaia} sources found in the field and the HST image to 20 sources. 5 of these sources are found in common between both telescopes. We are unable to align the F2100W image directly to \textit{Gaia} given that there are only three \textit{Gaia} sources detected in this image. Therefore, we align F2100W directly to the F770W image.

We estimate the uncertainty in our alignment by measuring the centroid of all available \textit{Gaia} sources in the aligned images and comparing these to the reference ICRS \textit{Gaia} coordinates. We find an error in our alignment of $\sim 32$ mas in the HST image and $\sim 17 - 22$ mas for most JWST images, except F2100W, which has the largest error of $\sim 37$ mas. In Figure~\ref{fig:target} we show the four JWST images and the HST image of SN\,2017gci. The black solid circle and cross-hairs show the location of the SN with coordinates from the GSA detections. The radius of the circle indicates our astrometric error, less than 1 pixel for all images. The red cross-hairs show the location of an extended emission feature seen in HST. The large black circle shows the aperture used to perform the photometry.

\begin{figure}
	\begin{center}
		\includegraphics[width=\columnwidth]{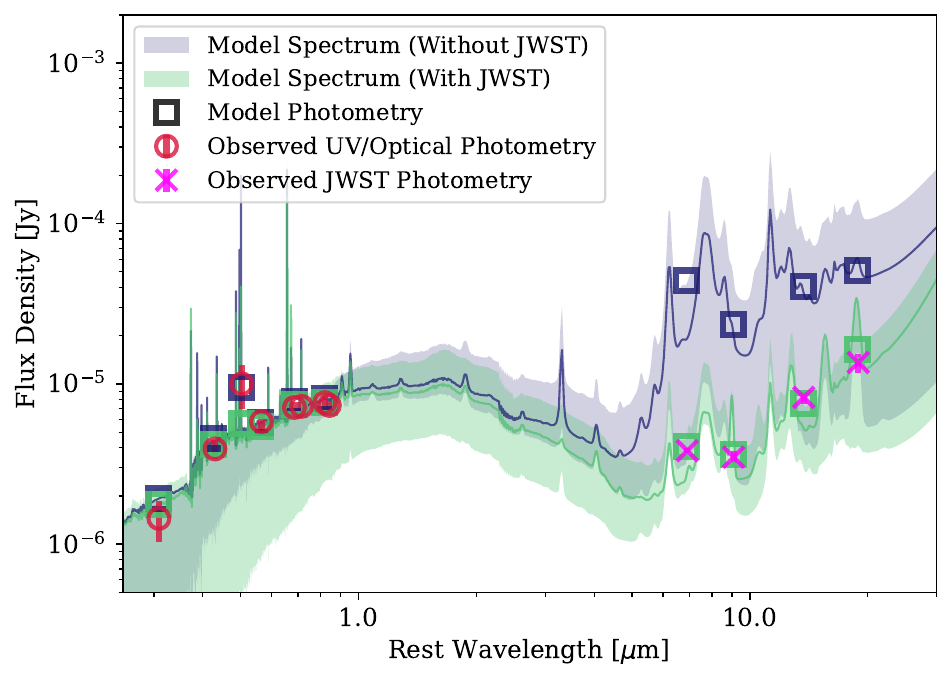}
		\caption{UV and optical photometry of the host galaxy of SN\,2017gci is shown as red circles, and JWST photometry as pink crosses. The blue line and shaded region represent the best-fit {\tt Prospector} model of the UV and optical photometry, without including the JWST photometry. The green line and shaded region represent the best-fit model that includes the JWST photometry. These models suggest that the JWST detections are most likely due to the host galaxy of SN\,2017gci, as opposed to the SN itself. \label{fig:prospector}}
	\end{center}
\end{figure}

We find that the location of the JWST emission is more consistent with the location of the extended emission rather than with the location of the SN. This is particularly evident in the bluer bands with smaller point spread functions (PSFs). While it is most likely that the JWST emission is not related to the SN, we cannot rule out a complete lack of IR emission from the SN based on this analysis alone given the proximity of the SN to the extended emission in the host galaxy of $\sim 1$ pixel, or $\sim 0.11$\arcsec. Therefore, we model the expected luminosity of the host galaxy in the following section.

\subsection{Host Galaxy Models}\label{sec:host}

We model the UV and optical photometry of the host galaxy of SN\,2017gci to quantify its expected luminosity in the JWST bands using {\tt Prospector} \citep{leja17, Johnson21}. We fit the spectrum using a delay-$\tau$ star formation history (SFH) with an exponential form of the form SFR $ \propto t^\alpha e ^{-t/\tau}$ \citep{Leja19}. We include the HST, PS1/$3\pi$, and DECam measurements, which are confidently lacking in SN flux. The resulting SED and best-fit {\tt Prospector} models are shown in Figure~\ref{fig:prospector}. We find that the expected flux of the host galaxy in JWST wavelengths is above the flux we measure in these images. This further supports the hypothesis that the JWST detections are due to the host galaxy of SN\,2017gci, and not the SN.

Additionally, we model the SED of the host galaxy in the same way as above, but this time including the JWST detections. We find an age of the galaxy of $3.4 \pm 1.1$ Gyr, a total mass formed of $\log(M / M_\odot) = 7.9 \pm 0.1$, a stellar metallicity of $\log(Z/Z_\odot) = -0.5 \pm 0.3$, a delayed SFH timescale $\tau = 4.7^{+2.8}_{-2.2}$, a largely unconstrained gas phase metallicity of $\log(Z_{\rm gas}/Z_\odot) = -0.03^{+0.33}_{-0.63}$, and an abundance parameter of polycyclic aromatic hydrocarbons of $Q_{\rm PAH} = -0.71^{+0.34}_{-0.15}$; this last parameter defined in \cite{Draine21}. We also explore a non-parametric SFH model with six bins of star formation and find consistent, albeit less constrained, results. This analysis supports the conclusion that the JWST detections are due to the host galaxy of SN\,2017gci.

\section{Modeling}\label{sec:models}

\begin{figure*}
	\begin{center}
		\includegraphics[width=0.95\columnwidth]{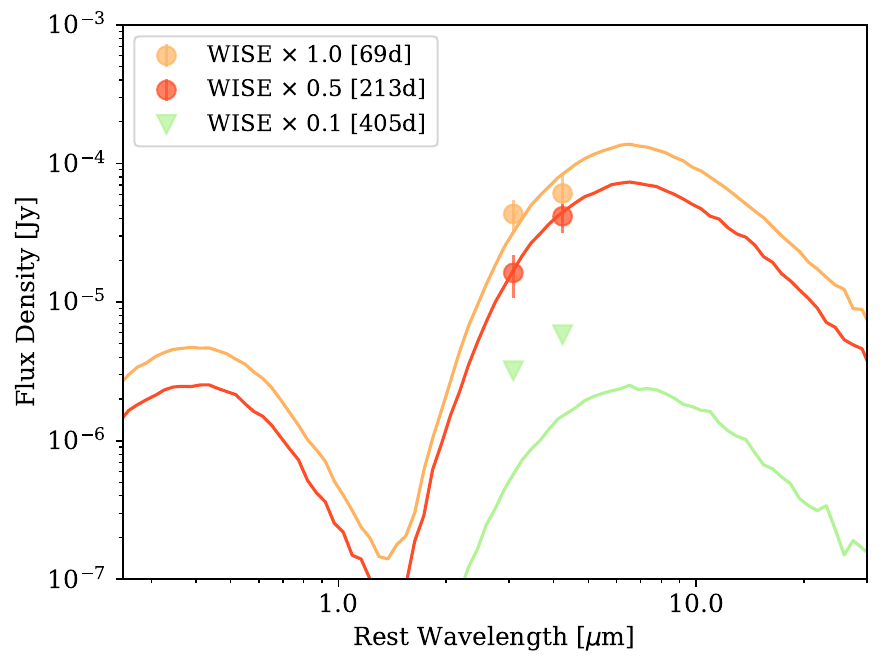}
		\includegraphics[width=0.95\columnwidth]{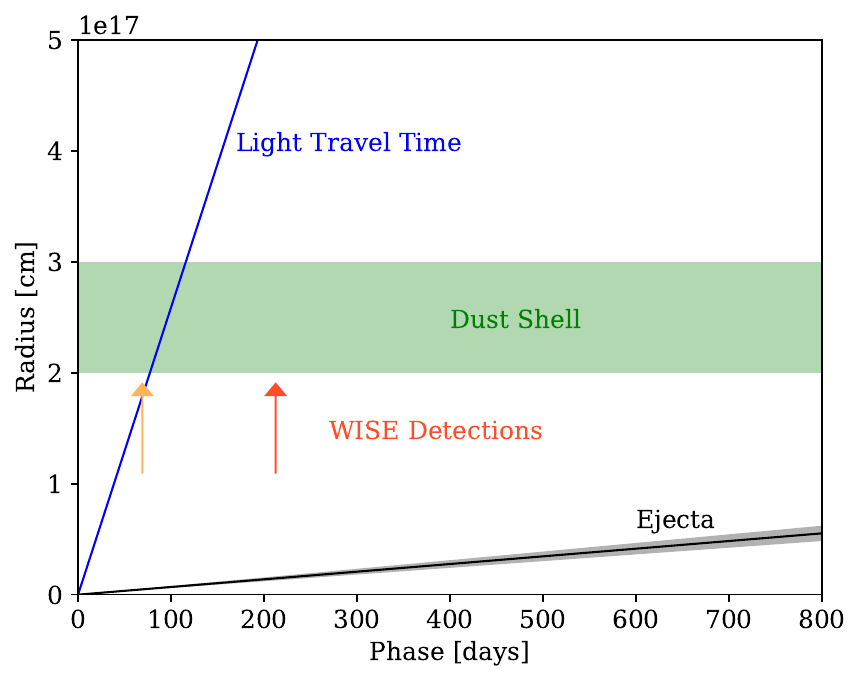}
		\caption{\textit{Left}: \WISE photometry of SN\,2017gci and corresponding {\tt MOCASSIN} dust models from a pre-existing dust shell. Distinct epochs are offset by 1 magnitude for clarity. We are unable to explain the non-detection at 405 days with freshly-formed dust. \textit{Right}: Diagram showing the location of the dust shell with respect to the ejecta. The \WISE detections come at a time that matches the light travel time to the inferred dust shell, suggesting the \WISE emission comes from pre-existing dust and not freshly-formed dust. \label{fig:wise}}
	\end{center}
\end{figure*}

\subsection{Pre-Existing Dust}\label{sec:wise_dust}

In Figure~\ref{fig:wise} we show the existing \WISE detections of SN\,2017gci at phases of 69 and 213 days, and subsequent non-detections at a phase of 405 days, originally presented in \cite{Sun22}. IR emission can usually be explained by either an echo from a pre-existing dust shell, interaction of the ejecta with CSM, or emission from freshly-formed dust.

We model the \WISE observations using {\tt MOCASSIN}, a radiative transfer code designed to model dust emission \citep{Ercolano03, Ercolano05}. We generate dust models based on the same framework used in \cite{Chen21} to model the dust emission of SN\,2018bsz. These models rely on the formalism described in \cite{Wesson10}, in which dust emission is calculated from within ellipsoidal regions bounded by the light-travel time. We assume dust emission from a spherical shell of CSM that surrounds the SN with a density profile proportional to $r^{-2}$. We reproduce the time evolution of the dust emission by assuming a thin radiation front that extends up to the light-travel time of the phase being modelled. Given that there are only two bands of photometry available, we are unable to constrain various parameters of the dust model and therefore assume a pure carbonate composition with grain radii of $a = 0.1\ \mu m$. We begin by assuming a heating source with a luminosity of $L = 1.0 \times 10^{41}$ erg s$^{-1}$ with a temperature of $T = 10,000$~K, based on the light curve properties of SN\,2017gci found by \cite{Fiore21}.

We vary the values of the luminosity of the heating source, the inner and outer radii, as well as the total mass to find a reasonable fit to the data. We find that a slightly brighter heating source of $L = 2.5 \times 10^{41}$ erg s$^{-1}$ heating a $M_{\rm shell} = 3.0 \times 10^{-4}$ M$_\odot$ shell of dust with an inner radius of $R_{\rm in} = 2.0 \times 10^{17}$ cm and an outer radius of $R_{\rm out} = 3.0 \times 10^{17}$ cm can reproduce the brightness and evolution of the SED well. In the left panel of Figure~\ref{fig:wise} we show the \WISE photometry and corresponding {\tt MOCASSIN} models.

In the right panel of Figure~\ref{fig:wise} we show the location of the dust shell with respect to the location of the ejecta. We assume an ejecta velocity of $V = 8000 \pm 1000$~km~s$^{-1}$ based on the measurements from \cite{Fiore21}. The fact that the dust shell lies well above the location of the ejecta supports the idea that the \WISE emission is coming from pre-existing dust. The velocity of the ejecta would have to be about an order of magnitude faster to reach the inferred dust shell location at the time of the \WISE detections.

We are unable to reproduce the increase in flux from 69 to 213 days and subsequent non-detection at day 405 with freshly-formed dust, since emission from freshly-formed dust would not be expected to fade quickly, but instead remain bright for longer, up to hundreds of days. Therefore, we conclude the \WISE detections are most likely due to a dust echo from pre-existing dust.

\subsection{Freshly-Formed Dust}\label{sec:jwst_dust}

In \S\ref{sec:methods} we concluded that the JWST detections are most likely due to the host galaxy of SN\,2017gci. Nevertheless, and for the sake of completeness, we explore the implication that these detections come from freshly-formed dust and derive upper limits on the total dust mass that could have formed.

\begin{figure*}
	\begin{center}
		\includegraphics[width=0.98\columnwidth]{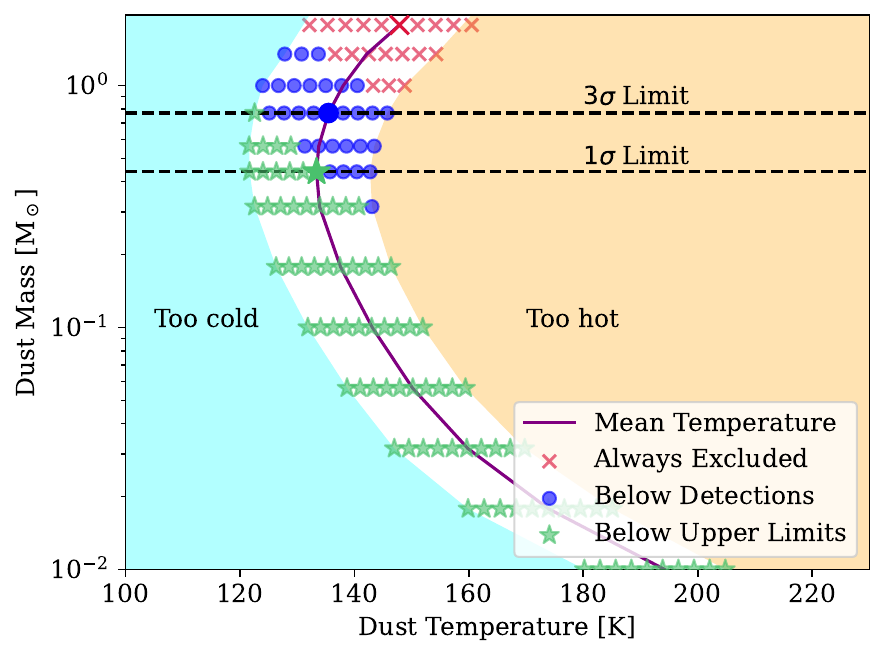}
		\includegraphics[width=1.10\columnwidth]{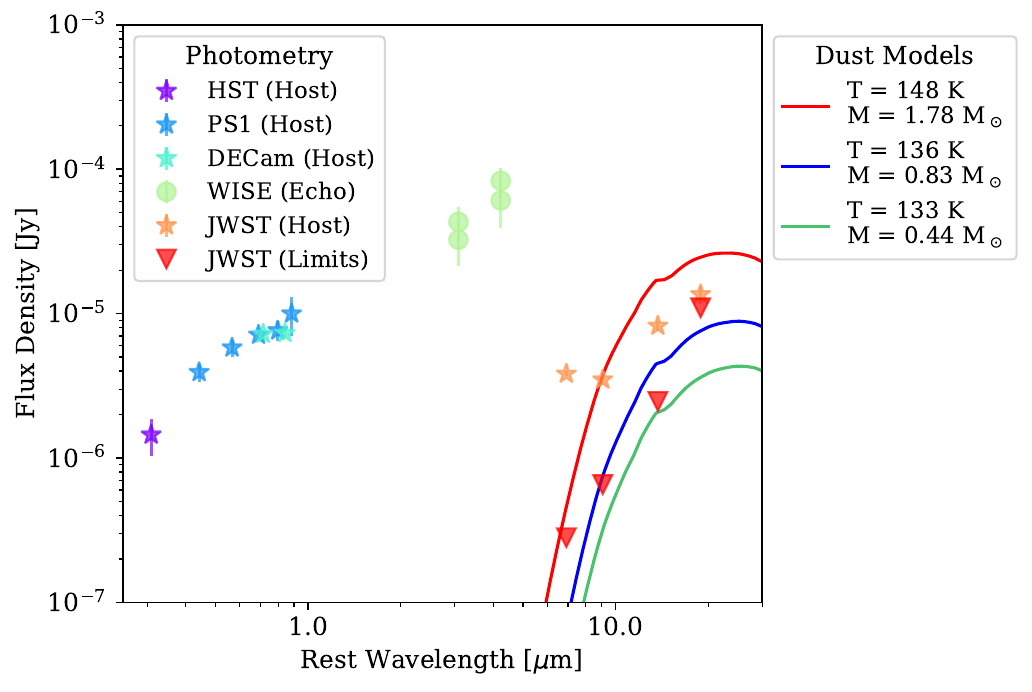}
		\caption{\textit{Left}: The cyan and orange regions are excluded based on a grid of heating sources modeled with {\tt MOCASSIN}. The green stars are models that cannot be excluded given that they lie below the upper limits. The blue stars are models that can only be excluded by the JWST host detections but still lie above the upper limits. The red crosses are models that lie above the detections and are therefore always excluded. The horizontal dashed lines the denote the 3 and 1$\sigma$ upper limits of 0.83 M$_\odot$ and 0.44 M$_\odot$, respectively. \textit{Right}: photometry of the field of SN\,2017gci, with photometry of either the host galaxy, the SN, or upper limits. The lines represent a selection of carbonate dust models, where the color of the lines on the right corresponds to the color of the models of the left. \label{fig:sed}}
	\end{center}
\end{figure*}

We estimate the temperature and luminosity of SN\,2017gci at the time of the JWST observations to simulate the source that would be heating any existing dust. We derive the luminosity of the SN by extrapolating the light curve models of \cite{Fiore21} out to 2252 days and find a most likely luminosity at the time of the JWST observations of $L = (1.2 \pm 0.38) \times 10^{39}$ erg s$^{-1}$. SN\,2017gci was observed to plateau at a temperature of $\sim 5000$ K at $\sim 150$ days after explosion \citep{Fiore21}. SLSNe have a temperature profile that tends to flatten as they evolve, remaining relatively constant for up to thousands of days after explosion. For example, the best studied SLSN, and the only one with a measurable temperature past 1000 days from explosion is SN\,2015bn, which plateaued at $\sim 7000$ K and cooled at a rate of only $68 \pm 23$ K every 100 days during phases of 400 to 1200 days \citep{Nicholl18_15bn}. If SN\,2017gci cooled at the same rate, we would expect the central heating source in SN\,2017gci to be $3590 \pm 530$ K at the time of the JWST observations. We note however that this is a lower estimate on the temperature of the heating source since the cooling rate likely slowed down with time. Additionally, any CSM interaction that might have occurred during the evolution of the SN could have further increased the temperature of the dust due to a reverse shock.

We adopt this range of luminosities and temperatures to simulate a heating source using {\tt MOCASSIN}, a code that performs the radiative transfer calculations necessary to derive the temperature of the dust shell. We explore an amorphous carbonate dust composition \citep{Hanner88} with dust grain radii of $a = 0.01\ \mu m$ or $a = 0.1\ \mu m$, and a shell of either constant density or a density profile of $r^{-2}$. Given the velocity of the ejecta of $\sim 8000$~km~s$^{-1}$, we explore a conservatively wide range of outer radii of $R_{\rm out} = 10^{16}$ to $2 \times 10^{17}$ cm, and an inner radius near the heating source of $10^{14}$ to $10^{15}$ cm. We find that the choice of dust radii, inner radius of the shell, and density profile have a negligible effect on the model. Therefore, we fix the grain radii to $a = 0.1\ \mu m$, the inner radius to be $10^{14}$ cm away from the heating source, and assume a constant density profile. We also fix the outer radius to $R_{\rm out} = 2 \times 10^{17}$ cm given that this choice produces the dimmest, and therefore most constraining, SEDs. We then derive a most likely dust temperature as a function of dust mass, with a corresponding $\pm 1\sigma$ range. The cyan and orange regions on the left panel of Figure~\ref{fig:sed} represent areas of parameter space that cannot be reproduced by any of these models and are therefore excluded.

We simulate a range of simpler dust models that span the allowed parameter space of temperatures and dust masses using a single component of carbonate dust. These models use the opacity parameters from \cite{Zubko04}, multiplied by a blackbody function scaled by the total dust mass. The rapid computation time of these models allows us to explore the entire allowed parameter space, and derive statistically meaningful limits on the dust mass. On the left panel of Figure~\ref{fig:sed} we show in red crosses the models that lie above the JWST detections, and are therefore always excluded. The blue circles are models that lie between the JWST detections and the inferred upper limits, meaning they are excluded only if we assume the JWST detections come entirely from the host galaxy. The green stars are models that lie below the upper limits, meaning we cannot reasonably exclude these models using the existing data.

In the right panel of Figure~\ref{fig:sed} we show three representative dust models for the most likely dust temperature at three representative dust mass values. In green we show the model with the lowest inferred dust temperature of $133$ K and a dust mass of $0.44$ M$_\odot$, which represents the lowest upper limit we derive. In blue we show a model with a dust temperature of $136$ K and a dust mass of $0.83$ M$_\odot$, a mass which can be excluded for 99\% of the allowed temperatures. In red we show a model with a relatively high dust temperature of $148$ K and a dust mass of $1.78$ M$_\odot$, which can be confidently ruled out for all allowed temperatures.

In conclusion, we can rule out with a $3\sigma$ confidence that more than $0.83$ M$_\odot$ of carbonate dust was formed in SN\,2017gci. Alternatively, our lowest estimate corresponds to a $1\sigma$ upper limit of $0.44$ M$_\odot$ of carbonate dust formed. These constraints are imposed by the highest possible SN flux derived from the upper limits shown as red triangles on the right panel of Figure~\ref{fig:sed}.

\section{Discussion}\label{sec:discussion}

\begin{figure}
	\begin{center}
		\includegraphics[width=\columnwidth]{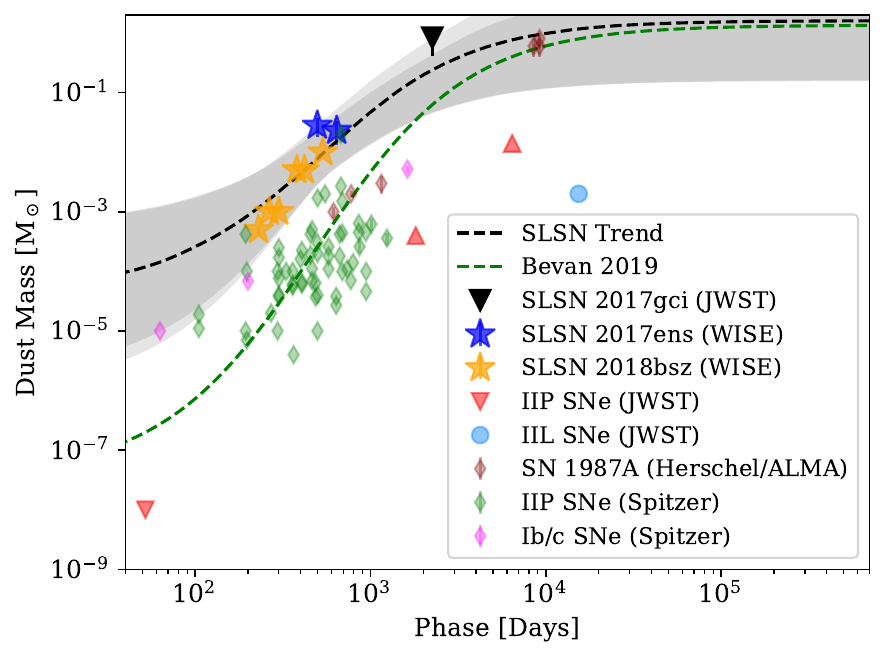}
		\caption{Measured dust masses for a range of SNe, including the SLSNe 2018bsz and 2017ens. The green line shows the nominal dust formation curve as a function of phase for normal CCSNe from \cite{Bevan19}. SLSNe with measured dust masses lie well above this trend line, $\sim 50\times$ higher. The black triangle represents the $3\sigma$ upper limit with an error bar extending down to the $1\sigma$ limit for the total mass formed in SN\,2017gci. The black shaded region and line represent possible curves of dust growth for SLSNe after taking into account the limit of SN\,2017gci. This suggests that SLSNe might form dust in a more accelerated fashion than other types of SNe. Original measurements from: \cite{Sugerman06, Meikle07, Kotak09, Sakon09, Andrews10, Andrews11, Fabbri11, Matsuura11, Szalai11, Gallagher12, Szalai13, Indebetouw14, Matsuura2015, Wesson15, Andrews16, Meza19, Tinyanont19, Tinyanont19_17eaw, Chen21, Rho21, Sun22, Ravi23, Shahbandeh23, Shahbandeh24, Zsiros24}. \label{fig:dust_masses}}
	\end{center}
\end{figure}

Bright IR emission detected in SLSNe hundreds of days after explosion led to the hypothesis that SLSNe could be very efficient dust producers. This argument is based on their inferred dust masses being higher than those of other types of SNe at similar epochs \citep{Chen21, Sun22}. Nevertheless, this excess in IR emission could have alternative explanations. For example, if a SN has a high ejecta velocity and therefore short diffusion timescale, this could result in a more rapid appearance of dust emission due to a rapidly declining opacity \citep{Dwek19}. This is unlikely to be the case for SLSNe given that they tend to have long diffusion timescales \citep{Nicholl15}. Moreover, SN\,2017gci has an ejecta velocity at peak of $\sim 10,000$~km~s$^{-1}$ \citep{Fiore21}, which is not particularly high for SLSNe, and comparable to the velocities of normal Type IIP and Ib/c SNe (e.g., \citealt{Takats12, Szalai13, Konyves21, Gomez22_LSN}). Alternatively, measuring dust masses in wavelengths bluer than where dust emission peaks can result in an underestimation of the total amount of dust. This is also unlikely to be the source of the high dust masses estimated for SLSNe given that these were measured using \textit{WISE} photometry, which is bluer than all other instruments used to measure the dust masses of other SNe we compare to (Figure~\ref{fig:dust_masses}).

It is possible however that SLSNe do not produce a higher total amount of dust, but instead have accelerated dust formation at early times that plateaus at a modest final dust mass earlier than for other types of CCSNe. Models of magnetar-powered transients support this hypothesis, suggesting that magnetars in SLSNe could accelerate dust formation, since the continued energy injection from a magnetar can promote faster condensation of dust grains \citep{Omand19}.

Typical SLSNe are thought to come from low-metallicity progenitors with zero-age main sequence (ZAMS) masses in the range of $\sim 20 - 30$ M$_\odot$ \citep{Blanchard20, Gomez24}. The theoretical maximum dust yield that could be formed by a SN from a low-metallicity progenitor in this mass range based on the available refractory elements is $\sim 1$ M$_\odot$ of carbonates and no more than $\sim 1.5$ M$_\odot$ of total dust \citep{Marassi19}. Even if SLSNe have a 100\% condensation efficiency, the total amount of dust they could form is at most a factor of $\sim 2$ higher than the $0.6 - 0.8$ M$_\odot$ of dust inferred for SN\,1987A \citep{Matsuura2015, Wesson15}.

In Figure~\ref{fig:dust_masses} we show how the dust masses of SLSNe compare to other types of SNe. Based on SN\,2017ens and SN\,2018bsz alone, it would be unclear whether SLSNe show an increased dust production throughout their entire evolution, or just at early times. Including the constraints derived in this work for SN\,2017gci, the former appears more likely. We fit the measured dust masses of SLSNe constrained by the limit of SN\,2017gci using the same non-physical function as \cite{Bevan19} to show possible curves of growth for dust emission. We see that the total dust yield for SLSNe is not necessarily higher than for other SNe such as SN\,1987A. If the growing observed dust mass shown in the Figure~\ref{fig:dust_masses} is not due to opacity effects, it is then likely that SLSNe form dust at an accelerated rate compared to other CCSNe.

\section{Conclusion}\label{sec:conclusion}

We have presented JWST/MIRI observations of the SLSN\,2017gci to constrain the total amount of freshly-formed dust in this SN. We detected emission in JWST/MIRI images, but find this to be most consistent with the host galaxy of SN\,2017gci, as opposed to the SN. This conclusion is based on both the significant separation between the location of the IR emission and the SN coordinates and the expectation for the IR luminosity of the host galaxy. We then use upper limits derived from forced photometry at the location of the SN to constrain the total amount of dust that could have formed. Our main conclusions are as follows:

\begin{itemize}
    \item \WISE observations between 69 and 405 days are best explained by light echos off a shell of pre-existing dust.
    \item JWST detections are most likely coming from the host galaxy of SN\,2017gci.
    \item Based on the upper-limits derived from the JWST photometry, we find that no more than $0.83$ M$_\odot$ of C/O dust formed in SN\,2017gci, with a lowest $1\sigma$ limit on the formed dust mass of $0.44$ M$_\odot$.
\end{itemize}

SLSNe have been suggested to be very efficient dust producers given their high inferred dust masses when compared to other types of CCSNe at similar epochs. The limits derived here for the total amount of dust formed in SN\,2017gci suggest that SLSNe do not necessarily produce a higher total amount of dust. If we assume the growth in the inferred dust mass is not due to opacity effects, it appears likely that SLSNe show accelerated dust formation at early times which plateaus earlier than for other types of CCSNe. A larger sample of JWST observations of SLSNe, particularly at redder wavelengths than the ones covered here, will be the key to determine whether this is the case.

\acknowledgments
S.G. is supported by an STScI Postdoctoral Fellowship. This project is supported in part by the Transients Science @ Space Telescope group. I.D.L. has received funding from the European Research Council (ERC) under the European Union's Horizon 2020 research and innovation programme DustOrigin (ERC-2019-StG-851622), from the Belgian Science Policy Office (BELSPO) through the PRODEX project “JWST/MIRI Science exploitation” (C4000142239) and from the Flemish Fund for Scientific Research (FWO-Vlaanderen) through the research project G0A1523N. D.L. is supported by research grants (VIL16599, VIL54489) from VILLUM FONDEN. T.S. is supported by the NKFIH OTKA FK-134432 grant of the National Research, Development and Innovation (NRDI) Office of Hungary.

IRAF is written and supported by the National Optical Astronomy Observatories, operated by the Association of Universities for Research in Astronomy, Inc. under cooperative agreement with the National Science Foundation. Operation of the Pan-STARRS1 telescope is supported by the National Aeronautics and Space Administration under grant No. NNX12AR65G and grant No. NNX14AM74G issued through the NEO Observation Program. This work has made use of data from the European Space Agency (ESA) mission {\it Gaia} (\url{https://www.cosmos.esa.int/gaia}), processed by the {\it Gaia} Data Processing and Analysis Consortium (DPAC, \url{https://www.cosmos.esa.int/web/gaia/dpac/consortium}). Funding for the DPAC has been provided by national institutions, in particular the institutions participating in the {\it Gaia} Multilateral Agreement. This research has made use of NASA’s Astrophysics Data System. This research has made use of the NASA/IPAC Infrared Science Archive, which is funded by the National Aeronautics and Space Administration and operated by the California Institute of Technology. This publication makes use of data products from the Near-Earth Object Wide-field Infrared Survey Explorer (NEOWISE), which is a joint project of the Jet Propulsion Laboratory/California Institute of Technology and the University of Arizona. NEOWISE is funded by the National Aeronautics and Space Administration. 

\facilities{ADS, Gaia, PS1, WISE}
\software{Astropy \citep{astropy}, extinction \citep{Barbary16}, Matplotlib \citep{matplotlib}, emcee \citep{Foreman13}, JHAT \citep{Rest2023}, \citep{numpy}, PyRAF \citep{science12}, SAOImage DS9 \citep{Smithsonian00}, space\_phot (\url{https://github.com/jpierel14/space_phot}), corner \citep{foreman16}, MOCASSIN \citep{Ercolano03}, prospector \citep{leja17}}

\bibliography{references}

\end{document}